\documentclass[pra,preprint,showpacs,preprintnumbers,amsmath,amssymb]{revtex4}
\usepackage{graphicx}
\usepackage{dcolumn}
\usepackage{bm}
\usepackage{mathrsfs}

\newtheorem{Definition}{Definition}

\begin{document}

\title{Residual Entropy and Refinement of the Third-Law Expression}

\author{Koun Shirai}
\affiliation{%
Nanoscience and Nanotechnology Center, ISIR, Osaka University, 8-1 Mihogaoka, Ibaraki, Osaka 567-0047, Japan
}%

\begin{abstract}
Although the third law of thermodynamics was established almost a century ago, it is not yet universally considered to be a fundamental law of physics. A major problem is that there are many materials having nonzero entropies at $T=0$, {\it i.e.}, residual entropy. Amorphous materials and random alloy systems are well-known examples. A conventional view is that amorphous materials are not in thermodynamic equilibrium and must be exempted from the law. 
The recent development of material sciences has let to a variety of new materials. Some of them have ambiguous structures which do not fit the qualitative description of metastability.
The definition of order states, on which old enunciations are based, also becomes vague. The establishment of an unambiguous statement which does not depend on the material properties is required.
This paper provides a quantitative expression for the third law to meet this requirement.
The idea is to introduce the notion of (thermodynamic) class. 
Every system belongs to a class. Different classes are thermodynamically separated by special internal constraints which are quantified by the frozen coordinate, $\hat {r}$. A frozen coordinate is a state variable that is not commonly possessed between two classes.  
The third law is restated as all materials within a given class have a common origin for their entropy. For two systems belonging to different classes, the entropy origins are shifted by the difference in entropy $S(\Delta \hat {r})$ associated with the difference $\Delta \hat {r}$ between the two systems.
When the internal constraint which maintains their respective thermodynamic equilibria is removed, the entropy origin must be reconstructed in order to establish only one thermodynamic equilibrium. This process is irreversible and this irreversibility is observed as the residual entropy $S(\Delta \hat {r})$. 
On the basis of this refinement, the residual entropies of amorphous materials as well as the long-standing problem of mixed states can be treated on an equal footing. 

\end{abstract}


\maketitle



\section{Introduction}

The third law of thermodynamics is a strange law in the physics literature. Despite being discovered almost a century ago, it does not seem to have attained a position as a fundamental law of physics. 
Among textbooks, different tones in the manner of exposition can be seen that vary from affirmative to reluctant \cite{Fowler-Guggenheim,Wilson,Lewis-Randall,Fermi,Pippard, Wilks,Guggenheim, Buchdahl,Beattie,Hatsopoulos,Gyftopoulos}. Many authors describe the law in restrictive manners, while some authors have given negative opinions on its importance as a fundamental law \cite{Landsberg57,Landsberg78,Landsberg97, Hasse,Tolman34,Tisza,CAQ,Baierlein,Waldram}. 
Many describe the law in only a few sentences. 
It is not uncommon to find no description at all.
This confusing situation becomes apparent by noticing that many studies on the third law are still being published. These recent studies will be cited in due course.

The central problem that has been plaguing us is the existence of so-called residual entropy. The most popular expression of the third law is called Nernst (heat) theorem.

\noindent
{\bf Third law: Expression I (Nernst theorem)}
{\em The entropy of any system vanishes as the temperature approaches zero.}

\noindent
This was first given by Nernst and later by Planck. The original expression given by Nernst was a little different from this. However, the assertion of the Nernst heat theorem after all turns to Expression (I) \cite{Cropper88}. This expression is very brief, but because of its briefness the third law has suffered repeated attacks. Although many thermodynamic phenomena such as chemical reactions follow this law, there are exceptions. Some materials have nonzero entropies at $T=0$, the well-known examples being amorphous materials, random alloys, ice crystals, and some asymmetric diatomic molecular crystals. Rich of these examples are well documented in classic thermodynamics textbooks \cite{Fowler-Guggenheim,Wilson,Lewis-Randall}. To circumvent this difficulty, restricted enunciations have been devised, such as, the restriction of the law to only chemically pure materials.
However, with the development of material sciences, subtle and marginal materials which do not fit the previous enunciations have been discovered. This has made it necessary to find a good reason for the existence of these exceptions. 
A new material requires another reason, and finally the law comes down into a sort of empirical rule.

On the other hand, there is another expression for the third law, namely, the unattainability of absolute zero temperature.

\noindent
{\bf Third law: Expression II (unattainability of zero temperature)}
{\em It is impossible to cool any system to absolute zero temperature.}

\noindent
Contrary to Expression (I), there are no exceptions in this case. Thus it is natural to consider that, if we start from Expression (II), we can obtain a rigorous expression in place of Expression (I) that does not allow any exceptions \cite{comment1}.
If fact, this was exactly the approach employed by Fowler and Guggenheim to derive their expression for the third law \cite{Fowler-Guggenheim}. While their expression (given later) has a great advantage in removing unnecessary restrictions, there is a difficulty that the notions, such as frozen state, given by them are only qualitative. 
Among newly discovered materials, there are many materials which do not fit these qualitative notions to describe a fundamental physics law. 

Clearly, a more accurate expression for the third law is required. 
The lack of a quantitative expression for residual entropy prevents universal acceptance of the third law. The purpose of this paper is to provide a general expression for the third law to meet this requirement.
To this end, it is essential to clarify the meaning of frozen states in a quantitative manner.
This is established by introducing the novel notions of the frozen coordinate and thermodynamic class, which are extensions of the ideas given by Hatsopoulos and Keenan \cite{Hatsopoulos}.
Throughout this study, it is stressed that thermodynamic theory must be constructed in a closed manner within a purely phenomenological framework; the first and second laws of thermodynamics do not depend on the microscopic mechanisms of matter. The same spirit must be maintained when describing the third law. This, in turn, will render the obtained expression robust with no exceptions.

\subsection{Classical-mechanics analogy}
\label{sec:brief}
To give readers a perspective on this paper, the idea is first explained by using a model of classical mechanics.
\begin{figure}[htpb]
\centering
\includegraphics[width=.80 \textwidth]{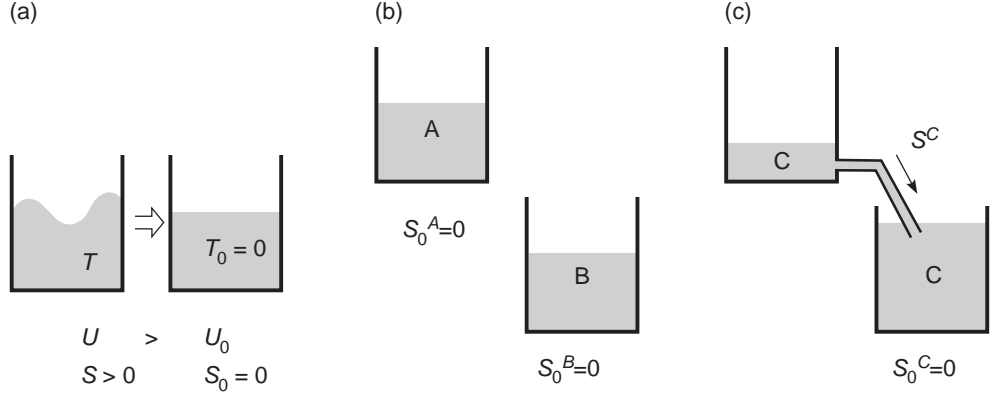}
\caption{
\textbf{(a)} State of water in a container. \textbf{(b)} Each volume of water having mass $M$ in two containers $A$ and $B$ is in the same lowest-energy state. \textbf{(c)} When the two containers are connected by a pipe, the combined system $C=A+B$ has a new origin of entropy.
}
\label{fig:separate-worlds} 
\end{figure}

\noindent
{\bf Example 1:}  
This is a model to illustrate the nature of residual entropy.
Water is contained in a container at ambient temperature $T_{0}$ (Fig.~\ref{fig:separate-worlds}(a)). 
In this model, when the water is at rest, the internal energy $U$ is the lowest $U_{0}$, and we take $T_{0}$ as our ``zero" temperature. The entropy $S$ is also zero, $S_{0}=0$. 
Do not consider the detailed behavior of fluids, and concentrate only on the energy balance, while the frictional energy is taken into consideration in order to reflect the thermodynamic aspect of the problem. 
When a disturbance is created in the water, the water enters in a higher-energy state, and $T$ and $S$ become to have finite values. 
Suppose that two containers are filled with water of the same amount (mass $M$), and are held with a difference in height $H$ (Fig.~\ref{fig:separate-worlds}(b)). Let denote the water in the higher container as $A$ and the water in the lower one as $B$. As long as the properties of water in the container are measured, there is no difference in the properties between water $A$ and $B$. The entropy origin of one system can be taken independently from that of the other.
The coordinate $H$ is irrelevant for describing the properties of the water in each container. The walls of the containers, which we call the {\em internal constraints}, maintain a constant value $H$. The quantity $H$ is called a {\em frozen coordinate}. When the two containers are connected by a pipe, water $A$ flows off toward $B$ (Fig.~\ref{fig:separate-worlds}(c)). We can look upon this as the occurrence of a thermal interaction between $A$ and $B$, creating a combined system $C=A+B$. Now, the entropy origins cannot be taken independently, allowing only $C$ to have the entropy origin $S_{0}^{C}$.
The fact that the state of $A$ has an extra coordinate $H$ starts to make sense. 
$H$ carries an extra energy $MgH$ ($g$ is gravitational acceleration). For the water flow is so fast, the potential energy $MgH$ is adiabatically converted to the kinetic energy of water, and $C$ enters an excited state at a slightly high temperature $T'$. This is an irreversible process, and the entropy is increased by $\Delta S(H)=MgH/T'$ \cite{comment-a}.
The excited state of $C$ finally dumps in the rest state, which is the lowest-energy state of $C$ and hence $S_{0}^{C}=0$. We observe that the entropy of $A$ was $\Delta S(H)$, which is interpreted as the residual entropy of $A$. 
This value $\Delta S(H)$ does not depend on materials in the containers. Any material in a container can be classified by the height $H$ with respect to entropy. A group of materials at the same $H$ forms a {\em thermodynamic class}.
Residual entropy thus arises when there is a state variable which is not commonly possessed between two systems.

The traditional view for the origin of residual entropy in Example 1 is that water $A$ is in a nonequilibrium state. This example clearly shows how irrational the traditional view is.  
Unfortunately, once we consider the inside of materials, the situation becomes unclear owing to the complexity of materials. Inside materials the role of internal constraint is played by an energy barrier, which prevents atoms from moving. The energy barrier is not a presence-or-absence sort of thing, and the hight of barrier is often marginal. All these factors obscure the origin of the residual entropy.
Therefore, a quantitative description for the above mentioned notions is indispensable.

The remainder of this paper consists of the following sections.
In Sec.~\ref{sec:preliminary}, a preliminary discussion on the meaning of zero entropy is given so that readers having different backgrounds can begin from the same starting point.
In Sec.~\ref{sec:critics}, previous statements of the third law are critically examined and the problems underlying these statements are analyzed.
Section \ref{sec:refinement} is the main part of this paper, in which the expression for the third law is refined in an unambiguous manner.
The conclusion is given in the last section.

\section{Preliminary consideration}
\label{sec:preliminary}
Before the main part of this study, several expositions on elemental notions in thermodynamics, on which the present arguments are based, are given. Even though these notions are reasonably well known, the author considers that not all of them are universally recognized. 

\subsection{Meaning of zero entropy}
\label{sec:meaning0}
\subsubsection{Common origin}
\label{sec:originS0} 
The original form of the third law was stated as an entropy difference. (i) The entropy change $\Delta S$ of a material vanishes as the temperature approaches absolute zero, 
\begin{equation}
\lim_{T \rightarrow 0} \left( \frac{\partial S}{\partial T} \right)_{X}=0,
\label{eq:dsdtX}
\end{equation}
where $X$ represents arbitrary state variables other than $T$. This relationship is only a fraction of the implications of the third law. (ii) The entropy difference in different states $X$ also vanishes,
\begin{equation}
\lim_{T \rightarrow 0} \left( \frac{\partial S}{\partial X} \right)_{T}=0.
\label{eq:dsdX-T}
\end{equation}
For example, the difference in entropy between different volumes $V$ for a given material becomes zero as $T \rightarrow 0$.
This implies that the entropy of a given material converges to a common value $S_{0}$. 
(iii) This common value is also shared by different phases of the same material. 
All crystals have more than one structures. At a certain $T$ and pressure, a crystal $A$ undergoes a phase transition from $\alpha$ to $\beta$, denoted by $\alpha \rightarrow \beta$.
The origins of the entropy of the two phases $\alpha$ and $\beta$ match each other,
\begin{equation}
S_{0}^{\alpha}=S_{0}^{\beta}.
\label{eq:comonab}
\end{equation}
The relation between diamond and graphite is an example of this. 
(iv) The implication of the third law is not restricted to within a given material but can be extended to different materials. In any chemical reaction $\alpha \rightarrow \beta$, Eq.~(\ref{eq:comonab}) holds. Here $\alpha$ and $\beta$ stand for the reactants and the reaction products, respectively, in a collective form. These may comprise several species. All the chemical species participating in this reaction share the same origin. For example, in the reaction ${\rm (1/2) C_{2} + O_{2}} \rightarrow {\rm CO_{2}}$, the three molecules, C$_{2}$, O$_{2}$, and ${\rm CO_{2}}$, have the same $S_{0}$. 
None of the above relations [Eqs.~(\ref{eq:dsdtX}) to (\ref{eq:comonab})] have an absolute value of entropy, and only differences appear. This implies that there is a common origin for different states and different materials. Expression (I) merely states that we have set the origin $S_{0}$ to zero. The absolute value of the origin is irrelevant \cite{Fermi,Wilson,Callen,Buchdahl}.


\subsubsection{Degeneracy}
\label{sec:degeneracy}
In quantum mechanics, there are many examples of systems having degeneracy. An electron with angular momentum $l$ has $(2 l + 1)$-fold degeneracy. Calculation by statistical mechanics gives $S_{0} = k_{\rm B} \ln (2l+1)$, where $k_{\rm B}$ is Boltzmann's constant. This seems not to cause any difficulty in thermodynamics. 
Similar situations also occur in classical mechanics. A rigid ball can move freely on the surface of a smooth table. The position ${\mathbf r}$ of the ball does not enter the thermodynamic description of the ball.
Hatsopoulos and Keenan termed this state as the {\it neutral state} \cite{Hatsopoulos}.
The states of a system are uniquely specified up to the freedom of the neutral states. 

When the same-energy states $\{ A_{i} \}$ of a system $A$ are separated by energy barriers, we can distinguish them as different states having different values of $X$.
In this case, we are vulnerable to falling in endless discussion of non-ergodic problem \cite{Lebowitz93,Barnum94,Jaynes65}.
In the macroscopic theory, however, there is no problem. At $T=0$, the barrier height, however small, is virtually equivalent to infinite. There is no possibility of a transitions from one state to another. Any two systems without thermodynamic communication are regarded as isolated systems. Each of the separated system has its own non-degenerate ground state. 

The relation of degeneracy to residual entropy is contemplated from a viewpoint of statistical mechanics.
The subtle relation between the thermodynamic limit and the temperature limit gives rise to the conditions for the emergence of residual entropy, which explain the microscopic origin of residual entropy \cite{Casimir63,Haar-Thermostat,Griffiths65, Leff70,Aizenman81,Chow87}. Later studies treated the residual entropy based on this theory \cite{Belgiorno03, Belgiorno03a, Masanes17}.
However, the issue for the macroscopic theory of thermodynamics is different. The issue is whether the presence of residual entropies indicates the violation of the third law or not. In the present study, we pursue to refine the statement of the third law in a manner consistent with residual entropy, while accepting its existence as experimental facts.


\subsubsection{Ordered states}
\label{sec:orderstate}
An easily misleading issue regarding entropy is the meaning of ordered state \cite{Ben-Naim}, which has an important consequence on residual entropy. This issue is a less-treated subject in standard textbooks. 

\begin{figure}[htpb]
\centering
\includegraphics[width=90 mm]{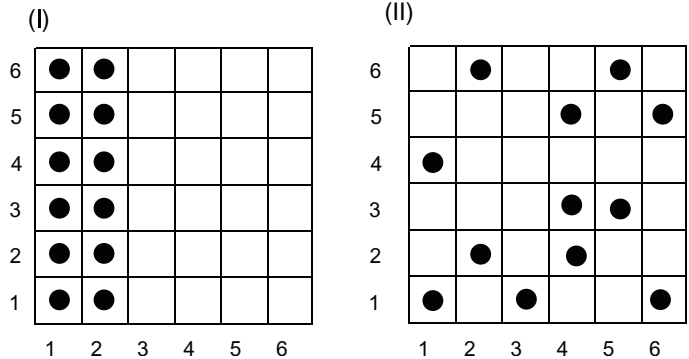}
\caption{
(I) ``Ordered" and (II) ``random" distributions of balls. Positions are specified by the row and column numbers $(m,n)$.
}
\label{fig:grid2} 
\end{figure}

\begin{figure}[htpb]
\centering
\includegraphics[width=90 mm]{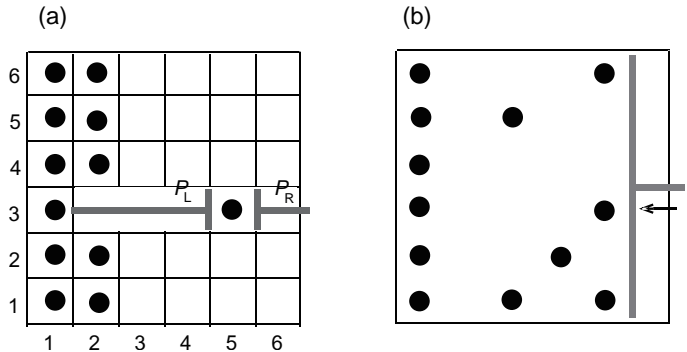}
\caption{
(a) Quasistatic way of moving a ball from $(3,2)$ to $(3,5)$. 
(b) Quasistatic way of recovering state (I) of Fig.~\ref{fig:grid2}.
}
\label{fig:grid3} 
\end{figure}

\noindent
{\bf Example 2:}
Figure \ref{fig:grid2} shows two arrangements of balls in a regular array of square walls. We consider which has the smaller entropy. In (I), all the balls are condensed on the left side in a closed pack manner, while in (II), the balls are distributed randomly. A reader may consider that arrangement (I) is the more highly ordered state with the lowest possible entropy. The fact is that the arrangement (I) and (II) have the same entropy of zero, as long as all the positions of the balls are known. If all the positions are known, we can move all the balls from arrangement (I) to (II) in a reversible manner without causing any effect outside the system. We will prove this by referring to Fig.~\ref{fig:grid3}(a).

System $A$ of Fig.~\ref{fig:grid3} is composed of many balls and the regular array of rigid walls filled with an ideal gas. Consider to move balls by two pistons $P_{\rm L}$ and $P_{\rm R}$.
The balls have no mass, and thus the work energy comes from compression/expansion of the gas solely. 
System $A$ has its entropy $S_{A}$. $A$ is in contact with a heat reserver $R$ at $T$, whose entropy is $S_{R}$. 
Let us move ball $B$ at position $(3,2)$ to $(3,5)$. To execute the move, all the walls in the third row are removed and are quickly replaced by two pistons, $P_{\rm L}$ and $P_{\rm R}$. Initially, the position of $B$ is fixed by $P_{\rm L}$ from the left-hand side and by $P_{\rm R}$ from the right-hand side. Let us quasistatically move $P_{\rm R}$ from $(3,3)$ to $(3,6)$. This movement of $P_{\rm R}$ is a reversible isothermal expansion of the ideal gas bounded by $P_{\rm L}$ and $P_{\rm R}$. Heat $Q$ is absorbed from $R$, and work $W$ with the same amount as $Q$ is transferred to an external device. Next, using this work $W$, we compress $P_{\rm L}$ in a quasistatic manner until it reaches position $(3,4)$. Then, $B$ is confined at the desired position. At the end of this process, $R$ recovers its original state, resulting in $\Delta S_{R}=0$. 
The whole process was carried out reversibly, meaning that $\Delta S_{A}+\Delta S_{R}=0$. 
We have achieved $\Delta S_{A}=0$ without causing any effect outside $A$. We can continue this procedure until configuration (II) is reached. 
We conclude that configurations (I) and (II) have the same entropy.

Our intuitive perception of randomness in arrangement (II) is based on the fact that we indeed do not know the positions of the balls in detail. 
This gives a configurational entropy as
\begin{equation}
S_{\rm conf} = -k_{\rm B} \left\{ c \ln c + (1-c) \ln (1-c) \right\},
\label{eq:confS}
\end{equation}
where $c$ is the ratio of the number of the balls to the total number of available cells.
Even in this case, by some means, we can restore the initial ordered state (I) from the disordered state (II) in a reversible manner. 
We can do this by sweeping quasistatically the whole space using a single wall from the right side (Fig.~\ref{fig:grid3}(b)).
However, more work is required than that when all the detailed positions were known. This extra work results in heat release to $R$. The increase $\Delta S_{R}$ is the unavoidable consequence of the fact that $A$ has a nonzero entropy $S_{A}$ in state (b).

The entropy increase is intimately associated with the notion of a {\it lack of detailed information} or {\it missing information} \cite{Haar-Thermostat,Ben-Naim, Jaynes57,Tribus71,Wehrl78,Zurek99,Leff}. 
For an extreme matter of black holes, missing information even gives a guiding principle to determine its entropy \cite{Bekenstein80,Dvali15}.
Apparent randomness is irrelevant to entropy.
To the human eye, the arrangement of nucleotides of DNA seems to have no regularity. Yet, its particular arrangement exactly conveys all the information about a complicated biological system. The entropy of DNA must be very low, if not zero.
%

The missing information in a transition $\alpha \rightarrow \beta$ can be quantified by the mapping of states.
There are numbers $W_{\alpha}$ and $W_{\beta}$ of the states of systems $\alpha$ and $\beta$, respectively. When $W_{\alpha} = W_{\beta}$, there is a one-to-one correspondence of states between $\alpha$ and $\beta$, and hence there is no missing information \cite{Bennett82}. This secures that the transition $\alpha \rightarrow \beta$ can be performed reversibly. When $W_{\alpha} < W_{\beta}$, the correspondence between two sets of states is one-to-many, and missing information occurs. This interpretation is consonant with the microscopic definition of entropy, $S=k_{\rm B} \ln W$, as should be.


\subsection{Other relevant issues}
\label{sec:existenceRev}
\subsubsection{Existence of reversible path}
Another issue which should not be overlooked in discussing the residual entropy problem is the existence postulation of a reversible path. 
Students learn that, in the second law of thermodynamics, the entropy difference $\Delta S_{21}$ between states 1 and 2 is defined as
\begin{equation}
\Delta S_{21} = \int_{1}^{2} \left( \frac{dQ}{T} \right)_{\rm (rev)}
\label{eq:defS}
\end{equation}
along a reversible path. On the other hand, it is rare to find, in textbooks, the description that a reversible path can always be found for {\it any} change of states. 
For mixing of two solutions, which is normally an irreversible process, it is known that the two end states can be reversibly reached by using semipermeable membranes. However, for solids, it is not at all evident. It seems difficult to recover the original crystal state from an ill-condensed structure by a reversible path. Nobody knows how to bring a dead body back to a living state. As far as the author knows, there is no proof for the existence of a reversible path, and we have to accept this as a postulation in thermodynamics.

\noindent
{\bf Existence postulation of reversible path}
{\em For any change between two states, there is always a reversible path connecting these two states.}

\noindent

\subsubsection{Axiomatic approaches}
Theorists have applied the axiomatic approaches to the third law \cite{Landsberg57,Giles,Lieb99, Wreszinski09,Belgiorno03,Belgiorno03a}. 
It seems that these axiomatic approaches do not fully treat the problem of residual entropy. These approaches mainly focus on the analytic behavior of the entropy function $S(\{ X_{i} \})$ as $T \rightarrow 0$, such as the continuity of the entropy function, boundary problem, and so forth.
The argument in Sec.~\ref{sec:frozen-state} shows that residual entropy emerges rather from where there is missing variables in the set $\{ X_{i} \}$ in order completely to specify the states of a given system. For solids, it is not obvious how to choose a complete set of variables, and this will be considered in Sec.~\ref{sec:example-third}.

\section{Critical consideration of previous expressions}
\label{sec:critics}

\subsection{Nonequilibrium character of frozen state}
\label{sec:FGstatement}
A widely held view on the residual entropy of glasses is that a glass is a metastable state. It is said that, in a geophysical time scale, a glass will be crystallized, and thus the third law will be recovered. This may be or may not be true: nobody can verify this. Our planet is filled with a tremendous number of metastable materials. All fuels are metastable, otherwise energy could not be extracted from them. H$_{2}$ and O$_{2}$ are metastable relative to H$_{2}$O. Diamond is metastable relative to graphite at ambient pressure. All these metastable states will be changed in a geophysical time scale. Nevertheless, diamond and graphite as well as many other materials commonly satisfy $S_{0}=0$. 
The metastability itself does not explain the residual entropy.

To distinguish the glass state from usual metastable states, Fowler and Guggenheim introduced the notion of a {\it frozen} metastable state (or simply frozen state); frozen states are essentially nonequilibrium states but are frozen by a strong viscous resistance \cite{Fowler-Guggenheim}. With using this term, Fowler and Guggenheim formulated the third law (\cite{Fowler-Guggenheim}, p.~226) as follows.
\begin{quote}
For any isothermal process involving only phases in internal equilibrium or, alternatively, if any phase is in frozen metastable equilibrium, provided the process does not disturb this frozen equilibrium,
\begin{equation}
\lim_{T \rightarrow 0} \Delta S = 0.
\nonumber
\end{equation}
(... )
Any conceivable isothermal process involving a phase in frozen metastable equilibrium, which does disturb the metastability, can obviously proceed only in the direction which decreases this metastability. 
\begin{equation}
\lim_{T \rightarrow 0} \Delta S < 0.  
\nonumber
\end{equation}
\end{quote}
This is abbreviated as the FG statement (see, however, \cite{comment2}).
The great advancement of the FG statement is that even for glasses we can take $S_{0}=0$ as long as the glass state retains its own structure. This message is reflected in Example 1.
Low-temperature experiments on glasses show that the specific heat converges zero as $T \rightarrow 0$, and the obtained value $S_{0}$ is independent of the pressure at which the experiment was performed. Only when the glass thaws, the entropy origin must be reconstructed in a similar manner to Example 1. This process is irreversible, and the entropy is increased by the residual entropy.

\begin{figure}[ht!]
\centering
\includegraphics[width=.35 \textwidth]{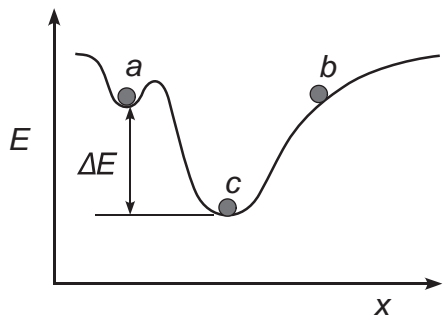}
\caption{
Distinction between metastable stable (a) and frozen state (b). (c) is the stable state.
}
\label{fig:internal-eq} 
\end{figure}

A problem with the FG statement is the ambiguity in the definition of the frozen state. 
The distinction between frozen and usual metastable states may be best seen in Fig.~\ref{fig:internal-eq}, which is taken from Wilks's textbook (\cite{Wilks}, p.~61). 
At first glance, the qualitative distinction shown in Fig.~\ref{fig:internal-eq} looks rational. However, if observing more closely, we will find no difference between metastable and frozen states. The atoms in a glass are moving around the average positions. The structure is no more than a local energy-minimum state, similar to usual metastable states. 
It is stable against small perturbations: The phonon spectrum does not show unstable modes \cite{Lucovsky}. In this respect, some ferroelectric materials are more unstable: They often exhibits phase transitions even at low temperatures. 
Accordingly, a more accurate definition of the frozen state is needed.

It is sometimes said that glasses are nonequilibrium states, simply because the properties of the obtained glasses depend on the process. However, this merely says that the obtained samples are different glasses having different properties.
It is an elemental leaning of thermodynamics that properties (or states) are different from processes. 
The glass transition is certainly nonequilibrium process and the obtained glasses will show different properties. This is not special to glasses but all transitions, by definition, are nonequilibrium processes. The mechanical strength of metals somehow depends on the heat history. The composition of a metal alloy obtained from a melt depends on the cooling rate. 
However, if we compare two samples having the same properties, we cannot distinguish the processes by which they were obtained. This is true for entropy too, because entropy is a state variable.

There is another problem in the second part of the FG statement. The second part implicitly premises that the lowest-energy state is the ordered state. For glass materials, this is certainly true.
However, the materials now to be considered are highly divergent, and among them many exceptions have been found. Frustrated systems in antiferromagnetic materials \cite{Magnetism}, spin glasses \cite{Magnetism}, geometrically frustrated systems \cite{Sadoc06}, incommensurate systems \cite{Janssen87}, quasicrystals \cite{Henley91}, are such examples. 
Among the many allotropes of boron, the lowest-energy structure of $\beta$-rhombohedral boron has a disorder structure if so defined, possessing a large residual entropy \cite{Ogitsu09}. The definition of the ``order" state is unclear when discussing the third law, as already pointed out in Sec.~\ref{sec:meaning0}. There is no fundamental reason in physics to prohibit a disordered state from being the lowest-energy state \cite{Wojciechowski91}.

\subsection{Irreversibility of transition}
\label{sec:BOstatement}
There is another viewpoint for treating the residual entropy problem. The viewpoint is that existence of residual entropy is ascribed to the irreversibility of the process in which the material is obtained.
Beattie and Oppenheim (the BO statement) state the following (\cite{Beattie}, p.~240).
\begin{quote}
Let a system B undergo the isothermal physical or chemical change in state
\begin{equation}
{\rm B(state \ 1)} = {\rm B(state \ 2)} \quad (T),
\nonumber
\end{equation}
where the initial and final states of B have the same temperature, but not necessarily the same values of pressure, volume, or any other property. If, as the temperature $T$ approaches 0 K, the above change in state may be brought about by any reversible process, which need not itself be isothermal, the third law requires that  
\begin{equation}
\lim_{T \rightarrow 0} \Delta S = 0. \quad 
\nonumber
\end{equation} 
\end{quote}
This says that if there is at least one reversible path from state 1 to 2, the two connected states have a common origin in $S$. In accordance with this, we can understand that gray tin and white tin have a common origin in $S$ owing to the fact that these two phases are connected by a reversible transition at a high temperature. Similarly, the fact that H$_{2}$ and O$_{2}$ have a common origin can be understood by considering that these two molecular states are reversibly transformed to H$_{2}$O molecule at a finite temperature. By controlling the concentrations of these molecules, we can perform the reaction in a reversible manner.
On the other hand, for glasses, it seems reasonable to conjecture as the origin of residual entropy that there is no reversible path from the glass state to the crystal state.

Using the BO statement, one is free from the ambiguity of the term ``metastability" in the FG statement. 
However, an inherent shortcoming of the BO statement lies in the use of processes for the judging residual entropy. When we interpret the BO statement literally, the statement is tantamount to that there must be no way to connect the state having residual entropy to the other states having no residual entropy in a reversible manner. However, in this case, discussing the entropy difference itself becomes meaningless because the entropy difference is obtained along a reversible path, as described in Sec.~\ref{sec:existenceRev}. Conversely, if we can find a reversible path, then this contradicts what the BO statement claims.
We have fallen into an impasse.
Beattie and Oppenheim theirselves were aware of this self-contradiction (\cite{Beattie}, p.~256). 
They seem to be content with giving a tentative conclusion that the calculation of thermodynamic quantities between two states which cannot be connected by any reversible path does not make sense. 
However, experimentalists indeed obtain entropy differences between glasses and crystals, and on the strength of this observation, we are discussing the mechanism of the residual entropy. 
Although the BO statement contains an important aspect of the third law, in this case, the expression should be revised by refining the meaning of ``no existence of reversible path", which will be shown in Sec.~\ref{sec:measure}. 

\subsection{Separation by internal constraints}
\label{sec:HKstatement}
Differently from the previous two statements, Hatsopoulos and Keenan gave a new statement for the third law (the HK statement) by introducing the new notion of a {\it quasistable state} (\cite{Hatsopoulos}, p.~568).
\begin{quote}
At zero temperature the entropy of any system is the same for all quasistable states of the system. Moreover, the entropy of any other state of the system at zero temperature is greater than that in a quasistable state.
\end{quote}
Quasistable states are defined by them as the states that can be made stable in a reversible manner merely by changing external agents, such as piston, external electric field.
These external agents are called external parameters $\beta$ \cite{Hatsopoulos, Gyftopoulos}. The external parameters are considered to be controllable.
The states that cannot be made stable in this way are those states which are separated from the quasistable states by internal constraints. These states are called {\it quasistatic} states. The issue of controllability/uncontrollability will be revisited in Secs.~\ref{sec:internal-constraints} and \ref{sec:measure}, while let us presently accept this definition.
The orientational disordered state of a diatomic molecular crystal is a quasistatic state, which is separated from the ordered state by the internal constraint. Glass states are quasistatic states, which are separated from the quasistable state of the crystalline form. In this manner, the internal constraint yields the residual entropy between the quasistatic and quasistable states.
From the HK statement, we can obtain a good criterion of whether a material has residual entropy or not, which depends upon neither the metastability of the material nor the irreversibility of the process in which the material was obtained. It depends entirely on the presence or absence of internal constraints. Accordingly, this statement has advantages over the previous two statements. 
We will develop a quantitative expression for the third law that is based on the HK statement in Sec.~\ref{sec:refinement}.

\subsection{Problem of mixing entropy}
\label{sec:mixE}

The residual entropy due to mixing entropy has long been a thorn in the side of thermodynamicists.
Mixing is observed in a wide range of phenomena, such as random alloys, impurity systems, isotopically mixed systems, and so on. The random orientations of electronic spins as well as nuclear spins also belong to this category. For this problem, maintaining the view that the state of nonequilibrium is the cause of residual entropy, which is traditionally adapted to glasses, is difficult.
It is absurd to consider that, if we wait for an astrophysical time, a random isotopic distribution will become an ordered one. Although some authors hold this extreme opinion, many ones keep silence on this problem.
Historically, because of this difficulty, Nernst and Planck restricted the validity of the third law to chemically pure systems only. Now, this restriction makes no sense. On the one hand, even a chemically pure crystal has a random distribution of isotopes. On the other hand, compound crystals, such as GaAs, are not chemically pure materials yet have no residual entropy. 

To circumvent the difficulty of isotopically random distribution, the idea of separating one property from other properties was devised by using the terms {\em aspect} \cite{Wilks,Atkins} or {\em factor} \cite{Simon37}. 
The mutual interaction between atom structure and isotope distribution is so weak that the two properties can be separately in thermal equilibrium.
Separation of the total entropy into the components is reasonable, and thereupon is later developed to a notion ``thermodynamic classes".
However, although separating material properties into different aspects secures the third law for the selected aspect, it does not save the remaining aspects from violation of the law. For example, the value $S_{0}=0$ of the perfect crystal is not threatened by the existence of random distribution of isotopes, if we separate the aspect of isotope distribution from the aspect of atom position. Despite this separation, the aspect of isotope distribution still have $S_{0} \neq 0$. From the latter fact, we can only deduced that the isotopically mixed state is in a nonequilibrium state. Thus, we have again encountered the original problem.

For a modest case of random alloys, some authors considered as the origin of the residual entropy that these alloys are in nonequilibrium states \cite{Simon37, Falk59}.
The Bragg--William model for the order/disorder transition \cite{Kittel5th} gives support for this view.
According to this model, at $T=0$, a composite system {\it AB} will be either a compound or separated to $A$ and $B$, depending on the sign of the bonding energy $V_{\rm AB}$ between $A$ and $B$ atoms. The random-alloy structure appears only at a finite $T$ due to the entropic effect.
The critical temperature $T_{c}$ of the order/disorder transition is given as $T_{c}=-2 V_{\rm AB}/k_{\rm B}$. 
However, the hypothesis that any mixed states are nonequilibrium states is not tenable. When $V_{\rm AB}=0$, this model asserts that the random alloys are stable even at $T=0$. It may be argued that this ideal case never occurs in the real world. However, physics laws hold better in the ideal case than in a real case.
Isotopically mixed crystals and random alloys are in any respect stable. In the macroscopic theory, thermodynamic equilibrium is so {\em defined}, if the state of a matter does not change in the time scale of interest, regardless of its microscopic structure. 
Problems of thermal equilibrium is further contemplated in Sec.~\ref{sec:internal-constraints}.

\section{Refinement of statement}
\label{sec:refinement}

\subsection{Internal constraints}
\label{sec:internal-constraints}
We are now ready to improve the expression of the third law.
On the basis of the foregoing arguments, we can lay the direction for the improvement. The keywords of the improvement are thermodynamic equilibrium, internal constraints, and frozen states. These words should be defined with no reference to microscopic mechanisms, and, if desirable, should be quantitatively defined.

The meaning of internal constraints is evident in macroscopic phenomena.
A following example is trivial.

\noindent
{\bf Example 3:} 
A rigid wall partitioning the inside a cylinder which is filled with a gas.

\noindent
In materials, the energy barrier is an incarnation of the abstract term ``internal constraint". 
An energy barrier prevents an unstable state from spontaneously collapsing into a stable state. 
Let us begin with investigating of the relation between thermodynamic equilibria and internal constraints.

A century ago, Gibbs introduced the notion of passive resistance in thermodynamics \cite{Gibbs}. His intention was to distinguish two categories in static states.
One case is static states which are achieved by the balance of the active tendencies of the system. There is another case of static states in which any change in the state is prevented by passive resistances. A mixture of N$_{2}$ and H$_{2}$ gases is stable at room temperature, even though a free-energy calculation shows that the reaction ${\rm N_{2} + 3 H_{2} } \rightarrow {\rm 2 NH_{3}}$ is expected. The energy barrier suppresses the reaction rate to such a low value that the reaction is almost completely inhibited. A passive resistance is a perfect inhibitor 
and is the same notion as an internal constraint. The static state of the latter type is called {\it quasistatic} state by Hatsopoulos and Keenan \cite{Hatsopoulos}.

Now, there is no reason to distinguish the above two categories of equilibria. 
Any system has internal constraints, whatever kinds. 
Even a simple N$_{2}$ gas in a box has internal constraints, for example, the energy barrier preventing the molecule from dissociation.
After all, we will find that every state variable has the respective internal constraint: the latter fixes the value of the former \cite{Reiss}. 
There is also no reason to distinguish the internal/external constraints by its controllability.
Amorphous materials can be, in some cases, changed to the crystalline form by applying a high pressure. See Sec.~\ref{sec:measure} further discussion.
The internal constraint is, according to Hatsopoulos {\it et al.}~\cite{Hatsopoulos,Gyftopoulos}, used for distinguishing {\it systems}.
The same materials but having different constraints are regarded as different systems. Ordered ice and disordered ice are different systems, which have their own equilibrium states.
\begin{Definition}[Thermodynamic equilibrium]
Any state of a system is called thermodynamic equilibrium, if the macroscopic properties of the system do not change during a time of interest.
\end{Definition}
Any microscopic information, such as order/disorder, chemically pure, is superfluous.
A frequently claimed assertion that, if we wait for a long time, disordered solids will become ordered states, is only an expectation. Rather, there are a pieces of evidence that glasses created more than millions years ago still remain unchanged \cite{Berthier16}, whereas no metal retains its initial state in this time scale, owing to corrosion and any other external influence. Physics laws should not depend on unproven speculations.


The author is aware that there are objections to this definition of thermodynamic equilibrium. In the literature of glass physics, glasses are commonly treated as nonequilibrium states.
From this viewpoint, theoretical studies on the residual entropy of glasses have been made \cite{Gujrati18, comment4}. 
Some authors use a special terminology of kinetic constraint \cite{Reiss,Kivelson99,Corti97}. These authors use it rather as artificial device in order to make thermodynamic analysis amenable. In the present study, the internal constraint is referred to as a real quantity, that is, an energy barrier inside materials.
Another parlance in the glass physics is the internal equilibrium \cite{Gujrati10,Gujrati18,Guggenheim}. This term is used to distinguish the glass states from normal equilibrium states. In this way, different authors use different definitions. Clearly, this situation is undesirable, and the standard usage should be established in near future.
Here, anyway, Definition 1 is adapted throughout the present study in the spirit of the classical thermodynamics. Needless to say, each law of thermodynamics must be consistent with other laws of thermodynamics.
The zero-th law of thermodynamics defines temperature in terms of thermodynamic equilibrium, no matter how complicated the internal structure of a material is. Temperature can, of course, be well defined for glasses too, as long as the thermodynamic equilibrium is defined as Definition 1. We have to retain this definition for the third law too. Consider a stone on a slope and presently at rest. If we wait for a long time, it may fall down. If we regard the state of the stone as nonequilibrium state only because of this expectation for the future, a sound theory cannot be constructed. 

\begin{figure}[htpb]
\centering
\includegraphics[width=.80 \textwidth]{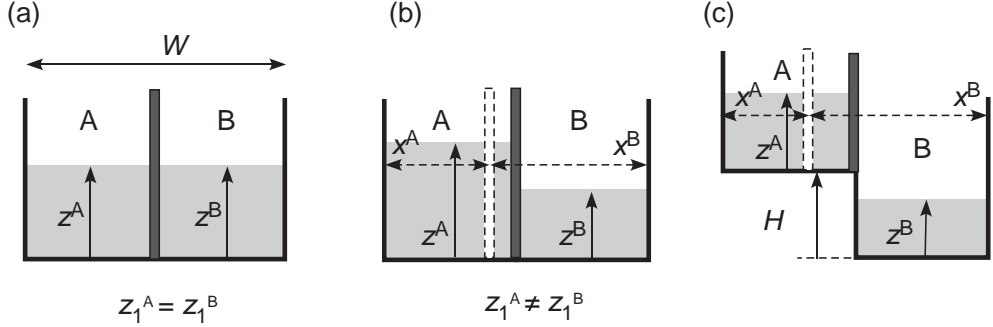}
\caption{
Two volumes of water ($A$ and $B$) partitioned by a wall in a container. \textbf{(a)} $A$ and $B$ have the same height. \textbf{(b)} $A$ and $B$ have different heights. \textbf{(c)} $A$ and $B$ are in different containers with their bottoms having a different height. In (b) and (c), the wall is mobile in the $x$ direction.
}
\label{fig:int-ext-constraints} 
\end{figure}

\subsection{Nature of frozen states}
\label{sec:frozen-state}
We have seen in Sec.~\ref{sec:HKstatement} that a residual entropy is sustained by an internal constraint.  Conversely, not all internal constraints necessarily carry residual entropies. A simple example is the fixed wall in Example 3. Removing the wall causes no residual entropy for both side gases before and after the removal of it. The gas remains the same gas before and after.
Let us analyze this difference in the internal constraints.
Figure \ref{fig:int-ext-constraints} illustrates situations similar to Example 1 with a slight modification. 

\noindent
{\bf Example 4:} The states of two volumes of water ($A$ and $B$) partitioned by a rigid wall in a container. 

\noindent
Three cases are discernible. 
In case (a), the initial heights of the two volumes of water, $z_{1}^{A}$ and $z_{1}^{B}$ are the same. Hence, after removing the wall, nothing happens. $A$ and $B$ remain in thermal equilibrium. 
In case (b), the initial heights, $z_{1}^{A}$ and $z_{1}^{B}$, are different. Removing the wall causes an irreversible process toward thermal equilibrium with the same heights $z_{2}^{A}=z_{2}^{B}$. We can still obtain the same result in a reversible manner by first moving the wall quasistatistically in the horizontal direction, $x^{A}$, until both heights become the same, and then removing the wall. 
Situations of chemical reactions are similar to this. 
Most chemical reactions $\alpha \rightarrow \beta$ are irreversible. However, by adjusting the temperature and/or concentration of molecules, chemical equilibria can be achieved.
No residual entropy emerges in case (b). A different situation occurs in (c). Thermal equilibrium cannot be obtained by only adjusting $x^{A}$. The difference $H$ in the depth of the bottoms of the volumes of water prevents $A$ and $B$ from reaching equilibria. In this case, a residual entropy $S_{0}$ associated with the new coordinate $H$ emerges. The advent of a new coordinate $H$, which we call a {\it frozen coordinate}, differentiates the frozen metastable state from usual metastable states.

Let us formulate the frozen coordinates quantitatively.
Henceforth, we use term {\it thermodynamic coordinates} for state variables \cite{Zemansky}, for the word coordinate is suitable for mathematical description.
The thermodynamic properties of a given system $A$ are described by thermodynamic coordinates $\mathbf q$; for example, for an ideal gas, $\mathbf q = (U, V, N)$, where $U$ is the internal energy of the gas and $N$ is the number of molecules in $A$.
Complete information of the thermodynamic properties of a system is given by the fundamental relation of thermodynamic equilibrium, $S=S(\mathbf q)$ \cite{Callen}. By using a minimum set of $m$ coordinates which are independent each other, $S$ is written as
\begin{equation}
S = S(q_{1}, \dots, q_{m}).
\label{eq:int-constraints-1a} 
\end{equation}
An important feature of thermal equilibrium is that all the thermodynamic coordinates are uniquely determined when thermodynamic equilibrium is established. This is a fundamental postulation \cite{Mackey89}: Hatsopoulos {\em et al} call this the stable-equilibrium-state principle \cite{Gyftopoulos, Hatsopoulos}.
When the total entropy $S$ can be factorized as
\begin{equation}
S(\mathbf{q}) = \sum_{j}^{m} s_{j}(q_{j}),
\label{eq:s-factorize}
\end{equation}
by applying an appropriate transformation among these coordinates, it is easy to see the nature of the entropy origin. Here, each component $s_{j}$ is a function of $q_{j}$ only, and hence it has the respective origin $s_{j,0}$. The origin of the total entropy hinges on the scope of the problem under consideration \cite{Jaynes65}. If we calculate the entropy of a given material by taking the spin freedom into account, the result will be different.

The values of the coordinates $\{ \mathbf{q} \}$ can be controlled by some means. By introducing an interaction between $A$ and $B$ through a coordinate $q_{j}$, the value $q_{j}^{A}$ begins to correlate to $q_{j}^{B}$. For example in Fig.~\ref{fig:int-ext-constraints}(b), $x^{B}=W-x^{A}$, where $W$ is the total width. From the knowledge of the one system, the state of the other system is known. This establishes a one-to-one correspondence between the two systems. The one-to-one correspondence implies no missing information as discussed in Sec.~\ref{sec:orderstate}. 
Having the same coordinates is not necessary for no missing information, whereas keeping a one-to-one transformation
\begin{equation}
q_{i}^{A} = f_{i}(q_{1}^{B}, \dots, q_{m}^{B}), \quad i=1, \dots, m,
\label{eq:int-constraints0} 
\end{equation}
is essential. For example, in Fig.~\ref{fig:int-ext-constraints}(b), a replacement of $z$ with $x$ does not destroy the one-to-one correspondence.

On the other hand, in case (c) of Example 4, a new coordinate $H$ appears. 
System $A$ has an extra coordinate $q_{m+1}=r$ ($H$ in this case), 
\begin{equation}
S_{A} = S_{A}(q_{1}^{A}, \dots, q_{m}^{A}, r^{A}).
\label{eq:int-constraints-1aa} 
\end{equation}
No one-to-one correspondence between two systems is possible.
\begin{Definition}[Frozen coordinates]
Frozen coordinate $r$ is a coordinate that is not common to two systems under consideration.
\end{Definition}
%
In order to remind that the coordinate $r$ is frozen, we denote it as $\hat{r}$, and hence $S_{A} = S_{A}(q_{1}^{A}, \dots, q_{m}^{A}, \hat{r}^{A})$.
As long as the internal constraint prevents $\hat{r}^{A}$ from varying, the thermodynamic properties of system $A$ do not depend on it.
In reality, it is not newly created in $A$, but is hidden in $B$ with a fixed value. Thus, we can conveniently take this fixed value as the origin of $\hat{r}$, namely, $\hat{r}^{B}=0$. 

Let us remove the constraint. The frozen coordinate $\hat{r}^{A}$ is activated to vary, and it becomes a real variable $r^{A}$. As seen in Eq.~(\ref{eq:int-constraints-1aa}), the dimensions of two systems are different. This breaks the one-to-one correspondence of Eq.~(\ref{eq:int-constraints0}). We say that there is missing information, resulting in an increase in entropy.
The relation between the one-to-one correspondence and the missing information was already mentioned in Sec.~\ref{sec:orderstate}.
This change $S_{0}^{AB}$ in entropy is discontinuous, because there is a jump in $r$ between $A$ and $B$. By using the factorized form Eq.~(\ref{eq:s-factorize}), $S_{0}^{AB}$ is expressed as,
\begin{equation}
S_{0}^{AB}=s_{A}(\hat{r}^{A})-s_{B}(\hat{r}^{B}).
\label{eq:res-r} 
\end{equation}
This is what we call the residual entropy $S_{0}^{A}$ of system $A$, when system $B$ is taken as the reference. 
By writing the total entropy in a form of Eq.~(\ref{eq:s-factorize}), we already saw that the value of entropy varies depending on which frozen coordinate is activated.
 
Buchdahl correctly figured out that the cause of residual entropy is the lack of uniqueness of the structure but not the metastability (\cite{Buchdahl}, p.~217). However, the recognition that this lack of uniqueness is caused by the discontinuity in the frozen coordinate was missing in his argument. 
Contrary, Hatsopoulos and Keenan were correct in recognition of the role of internal constraints for classification of static states \cite{Hatsopoulos}. 
However, they did not resolve the distinction between the two kinds of internal constraints: whether the removal of the internal constraint causes missing information or not.
In the literature of glass physics, special terminologies of the internal variable \cite{Bouchbinder09,Gujrati10}, the order parameters \cite{Moynihan81,Gupta76,Schmelzer06}, etc, are used. These terminologies may have a correspondence to the frozen coordinates in the present study. However, because the variety of terminologies are used in different contexts, it is difficult to categorize them. Reconciling these terminologies should be left as future work.

\subsection{Quantitative statement of the third law}
\label{sec:thermo-class}
Now we can introduce the new notion of thermodynamically different classes.
%
\begin{Definition}[Thermodynamic class of systems]
A thermodynamic class $\mathscr A$ is defined as a family of systems whose coordinates are connected by a one-to-one correspondence.
\end{Definition}
In stated differently, the idea of thermodynamic class is to classify thermodynamic systems according to the dimensionality of the system.
Two families whose coordinates do not have a one-to-one correspondence are called {\em thermodynamically different classes}. This is denoted by $\mathscr A \neq \mathscr B$.
Note that the same material can belong to different classes, depending to the scope of a problem. When the lattice contribution to the specific heat of a given crystal is considered, the atom positions $\{ {\mathbf r}_{j} \}$ are taken as the thermodynamic coordinates $\{ \mathbf{q} \}$. However, when the spin freedom is taken into account, this crystal belongs to a different class with additional coordinates of spins $\{ {\mathbf s}_{j} \}$ at each sites.

By using these terms, the third law can be stated in a quantitative manner.

\noindent
{\bf Third law: Expression III}

\begin{description}
\item{(i)} Within a thermodynamic class $\mathscr A$, every system has a common value for the origin of entropy as $T \rightarrow 0$.
\item{(ii)} For different thermodynamic classes $\mathscr A$ and $\mathscr B$, there is a difference $S_{0}^{\mathscr{AB}}$ in their origins of entropy.
The difference $S_{0}^{\mathscr{AB}}$ is associated with the discontinuity in the frozen coordinate $\hat{r}$ which separates one class from the other.
\end{description}

\noindent
This expression has no material dependence, no stable/metastable dependence, and no path dependence.
Expression (III) is free from the unproven hypothesis that disordered states are less stable than the ordered state. This is contrary to the second part of the FG and HK statements. Certainly, the entropy increases when the internal constraint is removed in accordance with the second law. However, it does not imply that the energy also increases. For isotopically mixed systems, a large entropy emerges, but virtually no energy gain is obtained by the mixing.

\section{Applications}
\label{sec:applications}

\subsection{Working examples}
\label{sec:example-third}
In thermodynamics, we are so accustomed to think that a great merit of thermodynamics is the extreme simplicity in describing systems, 
that we are liable to consider that the smallness of the number of coordinates is essential for describing the thermodynamic properties of systems. 
How small or how large the number of coordinates is irrelevant to the properties of thermodynamic equilibrium (\cite{Gyftopoulos}, Chap.~8).
For ideal gases, the fundamental relations are certainly expressed by only a few coordinates, {\it i.e.}, $S=S(U, V, N)$. For solids, which is virtually only a phase that the third law is relevant, the information of all the atom positions $\{ \mathbf{r}_{j}\}$$-$thermal averaged positions are meant throughout this paper$-$is needed to determine $S$. A gas state is the end state of losing the detailed information of atomic positions which the solid had. 
The general coordinates $\mathbf q$ for a crystal are given as $\mathbf q = (\mathbf{abc}; \{ \mathbf{r}_{\nu}\}_{\nu=1, \cdots s} )$, where $\mathbf{a},\mathbf{b},\mathbf{c}$ are the lattice vectors and $\mathbf{r}_{\nu}$ are the positions of atoms in a cell, in the case of $s$ basis atoms in the primitive unit cell. 
In the following, uninterested quantities $(\mathbf{abc})$ as well as $U$ are omitted from the coordinates.

For diamond and graphite, a one-to-one correspondence between the two coordinate systems holds; four atoms in two primitive unit cells of a diamond crystal ($s=2$) can be uniquely associated with four atoms in one primitive unit cell of a graphite crystal ($s=4$). There is no missing information, and hence the entropy origins are common between these two crystals.
For compounds, an additional label $\kappa$ of atomic species is needed in a form $\{ \mathbf{r}_{\nu}^{\kappa} \}$. However, atom species are uniquely associated with their sites $\mathbf{r}_{\nu}$, and hence $\kappa$ is a unique function of $\nu$, $\kappa=\kappa(\nu)$. 
Therefore, only $\{ \mathbf{r}_{\nu} \}$ are independent coordinates, which are sufficient to determine $S$.
In the formation of a SiC crystal with the diamond structure, we can uniquely map $({\rm \mathbf{r}_{1}^{Si}, \mathbf{r}_{2}^{Si}})$ of a primitive unit cell of a Si crystal and $({\rm \mathbf{r}_{1}^{C}, \mathbf{r}_{2}^{C}})$ of that of a diamond crystal to $({\rm \mathbf{r}_{1}^{Si}, \mathbf{r}_{2}^{C}; \ \mathbf{r}_{1}^{Si}, \mathbf{r}_{2}^{C}})$ in two primitive unit cells of a SiC crystal.
A one-to-one correspondence exists, and hence there is no difference in the entropy origins among these three crystals.
These examples are case (i) of Expression (III); these three crystals belong to the same class. 

\begin{figure}[htpb]
\centering
\includegraphics[width=.80 \textwidth]{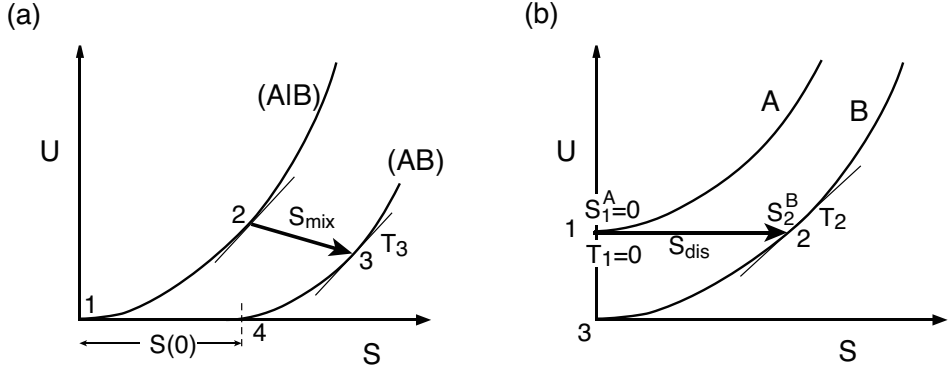}
\caption{
Heat paths crossing two thermodynamic classes. The processes removing the internal constraint are indicated by thick arrows. (a) A mixed system $(AB)$ is formed from the separated system $(A|B)$ in an isothermal process, $2 \rightarrow 3$. (b) The perfect crystal $B$ is recovered from a crystal $A$ having defects in an adiabatic process, $1 \rightarrow 2$. The process of reconstructing the entropy origin is always irreversible, in which $S$ increases. The gradient $dU/dS$ gives $T$.}
\label{fig:US-curves} 
\end{figure}

For random alloys, the periodicity of the Bravais lattice still holds. However, $\kappa$ is truly an independent coordinate, and in this case we need to know all the coordinates $\{ \mathbf{r}_{\nu}^{\kappa} \}$. 

\noindent
{\bf Example 5:} Silicon ($A$) and germanium ($B$) form a random binary system $A_{1-x}B_{x}$ with the same diamond structure, whose composition can vary over almost the entire range of $0 \leqq x \leqq 1$.

\noindent
The random crystal $A_{1-x}B_{x}$ is the mixed system, which is denoted by $(AB)$. The unmixed system is denoted by $(A|B)$. The internal constraint in this case is a spatial separation between the constituent crystals $A$ and $B$. 
In the mixed system $(AB)$, an atom position $\mathbf{r}_{j}$ is not associated with a particular atom species $\kappa$. 
There is no one-to-one correspondence in the atom positions between $(A|B)$ and $(AB)$; the two systems belong to different classes. There is missing information, resulting in a residual entropy given by Eq.~(\ref{eq:confS}). In this case, $\{ \kappa_{j} \}$ can be taken as the frozen coordinate $\hat{r}$. 

Now let us remove the internal constraint. This process is shown by path $2 \rightarrow3$ in Fig.~\ref{fig:US-curves}(a). When there are no appreciable energy differences among the interatomic bonds, the natural process occurring at a finite $T_{2}$ is the mixing of pure components. The process proceeds isothermally, and $S$ is increased by the mixing entropy $S_{\rm mix}$. Upon cooling from $T_{2} =T_{3}$, the mixed and unmixed crystals undergo changes in the $U-S$ curve in almost parallel ways because the bond energies are almost the same. Therefore, the entropy difference by the amount of $S_{\rm mix}$ remains down to $T=0$, which is observed as the residual entropy, $S_{0}=S_{\rm mix}$. We conclude that the frozen state is the minimum-entropy state, whereas the stable state is the higher-entropy state at $T=0$.

{\bf Example 6:} Electronics-quality silicon crystal is a crystal as perfect as humans have ever obtained, and thus there is no more ideal material than Si for studying zero entropy. Defects can be introduced in a Si crystal by electron irradiation at a low temperature. Let us denote a crystal with defects as $A$ and a perfect crystal $B$. A number of Si atoms have been kicked out into interstitial sites. Out of $N_{I}$ interstitial sites, $n_{I}$ of sites $\{ \mathbf{r}_{j}^{I} \}$ are occupied by Si atoms, creating a configuration entropy of $S_{\rm dis} = k_{\rm B} \ln(N_{I}/n_{I})$ due to disorder, which becomes the residual entropy $S_{0}$.
An energy barrier is built around an interstitial position $\mathbf{r}^{I}_{j}$, and thus the interstitial atom is fixed there. The positions $\{ \mathbf{r}_{j}^{I} \}$ can be regarded as the frozen coordinates.
In spite of the nonzero $S_{0}$, we can consider $A$ as having zero entropy $S_{1}^{A}=0$ at $T_{1}=0$ as long as $\{ \mathbf{r}_{j}^{I} \}$ does not change (see Fig.~\ref{fig:US-curves}(b)) consistent with the FG statement.
The defect state is thermodynamically stable at low temperatures.

These stable defects can be remedied by annealing at a high temperature $T_{a}$. The energy barrier can be surmounted by the thermal energy. To simplify analysis, suppose that the energy barrier can be removed at $T_{1}=0$ by some means. Also suppose that this was done adiabatically, so that the internal energy $U$ did not change.
The defect state has a higher energy than the energy of the perfect state. By removing the constraint, defect atoms move toward regular positions. The process is indicated by the thick arrow line $1 \rightarrow 2$ in Fig.~\ref{fig:US-curves}(b). The potential energy of a defect atom is converted to the kinetic energy. 
At the end of the process, the perfect crystal $B$ is recovered. However, it will be found that $B$ is at a finite temperature $T_{2}$ because of the adiabatic process.
The process of removing the internal constraint is the process by which the entropy origin is reconstructed from $0 \ (\equiv S_{1}^{A})$ to a finite value $S_{0} \ (\equiv S_{2}^{B})$ at $T_{2}$.
By cooling process $2 \rightarrow 3$, the lowest-energy state of $B$ with $S_{3}^{B}=0$ is obtained. We conclude that, in this case, the frozen state is a higher-entropy state at $T=0$.
The contrasting results of Examples 5 and 6 show that the second part of the FG and HK statements that the breaking of the frozen state always causes a decrease in entropy is not correct.

It is pertinent here to make a comment on the way of taking the entropy origin, by taking glass materials as an example. Although the glass case is very complicated, the scheme shown by Fig.~\ref{fig:US-curves}(b) is still valid.
We have already leaned that the glass state can be seen as a thermodynamics equilibrium state, which forms one thermodynamic class. As long as the glass belongs to this class, the residual entropy $S_{0}^{A}$ can be taken to be zero. 
Along the path $1 \rightarrow 2$, the internal constraint which maintained the glass state is removed. This process is very similar to the process of Example 1 (see Fig.~\ref{fig:separate-worlds}(c)). The potential energy maintained by the internal constraint is converted to the kinetic energy, and the system reaches state 2 of the crystal. This process associates an increase in $S$ from 0 to $S_{2}^{B}$. This increase is in accordance with the second law of thermodynamics.
Conversely, let us consider the cooling process $2 \rightarrow 1$.
There is an argument that the entropy is decreased from $S_{2}^{B}$ to 0; this decrease is not compensated by any heat generation; thus, the conclusion of this process is a decrease in the total entropy including the surroundings. This contradicts the second law. For this reason, the residual entropy of glasses was doubted by Kivelson {\it et al.} \cite{Kivelson99,Gupta07,Gupta08}; the measured value of $S_{0}$ should be an artifact arising from the irreversibility of the freezing process.
On the other hand, Goldstein showed that the nonzero residual entropy is real based on his analysis \cite{Goldstein08,Goldstein11}. See further discussions in a special issue of journal Entropy \cite{comment4,Proc-Trencin}. 

The contention of their debate lies in the way of taking the entropy origin between $A$ and $B$. The building of an internal constraint is a process in which one thermodynamic class is separated into two classes which have their own origins of entropy. When we need to evaluate the difference in $S$, somehow the common origin should be used. When the melting state of $B$ is supercooled to state 2, $S$ is decreased to $S_{2}^{B}$. This value is still as large as that of the liquid state.
After the glass transition is completed, all the atom positions are fixed. The atom arrangement of the glass is certainly only one among many arrangements. 
However, we do not know which arrangement actually occurs. 
There is missing information similarly to in Example 2, and the residual entropy is a real quantity. 
By observing from $B$, the residual entropy of $A$ is the large value $S_{2}^{B}$. This problem of glasses will be discussed in more details in a future study.


\subsection{Measurability}
\label{sec:measure}
Lastly, the measurability of the residual entropy is discussed.
There are two kinds of difficulties in experiments. The one is a practical problem of sample preparation. 
To obtain $S_{0}$, obviously we need both the ordered and disordered states.
For many cases, the freezing temperature is so high, such as the case of ice crystals, that it seems impossible to obtain the perfectly ordered crystal. The residual entropy can only be found by comparing the experimental value of the frozen state with the theoretical value of the hypothetical ordered state. Nonetheless, according to the existence postulation of the reversible path, there must be in principle a reversible process connecting the ordered and disordered states.
Useful methods of experiment, such as use of catalysis, were given in the literature \cite{Suga05, Kozliak08}.

The next problem involves a conceptual difficulty. In Example 5, the disordered crystal is obtained by mixing two pure crystals. Figure \ref{fig:US-curves}(a) shows this process (path $2 \rightarrow 3$).
The mixing process is essentially an irreversible one. 
At first glance, it seems to support the BO statement, namely, the irreversibility is the cause of the residual entropy. The Eastman and Milner's experiment on a solid solution of AgCl-AgBr is remarkable; first no reference of theoretical value was used, second no heat measurement was used, third no use of separation process by melting was made \cite{Eastman33}. The second method gives incorrect values of $S$ for irreversible processes, and the third method has a difficulty in reversible separation.
They employed an electrochemical method to obtain the mixing entropy $S_{\rm mix}$.
This measurement process corresponds to the path $3 \rightarrow 2$ in Fig.~\ref{fig:US-curves}(a) in a reversible manner. This means that we have found a reversible path from the mixed state 3 to the unmixed state 2.
The BO statement dictates zero $S_{0}$ when a reversible path is found. This conflicts with the experimental result of a nonzero $S_{0}$. We have fallen into a circular argument, which was encountered in Sec.~\ref{sec:FGstatement}.

The way of escaping from this circular argument is to understand carefully the meaning of the existence postulation of a reversible path.
By stating that a mixing process $A \rightarrow B$ is irreversible, it means that the process is a spontaneous change {\em within} the isolated system: in this case, the combined system $A+B$. In an isolated system, any spontaneous change is an irreversible process. However, this irreversible process can be replaced with a reversible one by bringing another system $C$. 
The electrochemical method is such a process. By connecting $A+B$ with an external voltage, an electric current flows. Behind this interaction, part of the mixed solid is melted back to the electrolyte. This establishes a reversible separation from the mixed state to its constituents.

This example gives us a guiding principle for measuring $S_{0}$. Connect $A$ and $B$ by an external agency $f$ in a manner that the resulting perturbation is minimum, retaining the initial states of $A$ and $B$ as much as possible. Such a method must always exist, by virtue of the existence postulation of reversible path. By knowing the relationship between $f$ and $S$, we can obtain $S_{0}$.



\section{Conclusion}
\label{sec:conclusion}

An unambiguous expression for the third law has been found.
Expression (III) is the most desirable one ever obtained, which meets all requirements that were pursued in the present study. No ambiguity as to whether a material is in metastable state or not enters. 
The irreversibility of the process in which the material is obtained does not enter.
It does not rely on the speculation that a disordered state will become eventually an ordered state, if we wait for a long time. 
It is free from the unproven hypothesis that the energy of any disordered state is higher than that of the ordered state.
The condition of controllability/uncontrollability of internal constraints is also rejected from the judgement of the residual entropy.

\begin{acknowledgments}
The author thanks F.~Belgiorno for useful criticisms on many aspects of the third law and Y.~Oono on the aspect of nonequilibrium states. From the material side, the author much learned quasicrystals from M.~Widom and K. Kimura, glass physics from O. Yamamuro, and Bose-Einstein condensation from J. C. Wheeler. Especially, the author got many benefits from the debate with P. D. Gujrati on the glass physics.
\end{acknowledgments}







\end{document}